# On the proficient use of GEV distribution: a case study of subtropical monsoon region in India

**Ripunjai K. Shukla**
**Department of Remote Sensing, B.I.T., Mesra, Ranchi**
ripu121@yahoo.co.in
**M. Trivedi, Manoj Kumar**
**Department of Applied Mathematics, B.I.T., Mesra, Ranchi**

**ABSTRACT:** The paper deals with the probabilistic estimates of extreme maximum rainfall (Annual basis) in the Ranchi, Jharkhand (India). Extreme Value Distribution family models are tried to capture the uncertainty of data and finally Generalized Extreme Value (GEV) distribution model is found as the best fitted distribution model. The GEV model satisfied the selection criteria [Anderson-Darling test (A-D test or Goodness of fit test) and Normality test (Q-Q plot)], which are adopted under the present study. The return levels are estimated for 5, 10, 50, 100 and 200 years which are consistently increasing for long run in future.

**Introduction**

Climate change is one of the great environmental concerns facing mankind in the twenty first century. The greatest threat to humans (and other components of terrestrial ecosystems) will be manifested locally via changes in regional extreme weather and climate events like draught due to high temperature, flood due to extreme rainfall. Extreme value theory deals with the stochastic behavior of the extreme values in a process. For a single process, the behavior of the maxima can be described by the three extreme value distributions – Gumbel, Frechet and negative Weibull – as suggested by [FT28]**.** [KN00] indicated that the extreme value distributions could be traced back to the work done by Bernoulli in 1709. The first application of extreme value distributions was probably made by Fuller in 1914.

An important issue in this research is to assess and predict the changes in extreme maximum rainfall in the future climate of Ranchi in Jharkhand, India. Jharkhand is a huge store of coal mining and it is badly affected by extreme rainfall. Rain water entered into mine areas and forcing the coal companies to stop mining and due to this production is badly affected at much extent. Several researchers have provided useful applications of extreme value distributions to rainfall data from different regions of the world: see [NNA02, NNW98] for applications in Canada**;** [KB00] for applications in Greece; [Fer93] and [Par99] for applications in India; [C+02] for applications in Italy; [EA93] for applications in Jordan**;** [Z+02] for applications in Malaysia; [WN00] for applications in New Zealand.

**Extremities in Weather**

Worldwide areas today severely affected by the extreme rainfall events. An accurate estimate of frequency and distribution of these events can significantly aid in policy planning and observation system design. Increased extreme precipitation events can be attributed to increase in moisture levels, thunderstorm activities and large scale storm activity. In the global warming scenario, climate models generally predict an increase in large precipitation events The numerical modeling community and data analysts have shown interest on the issue of extreme events occurring around the world.

Ranchi has a sub-tropical climate. During summers (March to June), the maximum temperature observed is 37°C and a minimum of 20°C. Winters (November to February) are cool and have a maximum temperature of 22°C and a minimum of 2°C. Monsoons (July to September) offer average rainfalls (http://www. Mustseeindia.com/Ranchi-weather). In the summer or winter, there is very little rainfall in Ranchi. However, in the monsoon season stretching from June to September, Ranchi experiences an average rainfall of 1,100 millimeters (http://www.mapsofindia.com/ ranchi/geography/ weather.html).

The objective of this study is to fit the efficient probabilistic approach from the family of extreme value distribution models, which can suitably model the uncertainties pattern in extreme maximum rainfall (mm) in subtropical monsoon region of India. Ranchi region is taken as key place to execute this study.

## 1. Methodology

The brief historical development of extreme value is given in the encyclopedia [KJR85] states that the basic theory of extreme value was first developed by Frechet in 1927 and by Fisher & Tippet in 1928 but was formalized by Gnedenko in 1943. Suppose there exists an independent and identically distributed (iid) sequence of random variables $X_1, X_2, \ldots, X_n$. Whose cumulative distribution function (CDF) is:

Also $M_n = Max(X_1, \ldots \ldots X_n)$ which is the $n^{th}$ sample maximum of the process and $M_n$ has CDF: $P(M_n \leq x) = [F(x)]^n$ ...............(1)

Equation (1) states that for any fixed x for which $F(x) < 1$, we have $Pr\{M \leq x\} \to 0$ as $n \to \infty$ which is not useful. However sequences of constants $a_n, b_n$ exist such that

$$P\left[\frac{M_n - b_n}{a_n} \leq x\right] = [H(a_n x + b_n)]^n \to H(x) \text{ is independent of n.}$$

According to extreme value theory $H(x)$ must be one of the three possible forms of distributions. The importance of this result is that irrespective of what the original distribution F is, the asymptotic distribution of $X_{(n)}$ is any of the three forms of the extreme value distribution. In their simplest form three types of the extreme value distribution are:

$$\text{I: } G(x) = \exp\left\{-\exp\left[-\left(\frac{x-b}{a}\right)\right]\right\}, \quad -\infty < x < \infty$$

$$\text{II: } G(x) = \begin{cases} 0, & x \leq b \\ \exp\left\{-\left(\frac{x-b}{a}\right)^{-\alpha}\right\}, & x > b \end{cases}$$

$$\text{III: } G(x) = \begin{cases} \exp\left\{-\left[-\left(\frac{x-b}{a}\right)^{\alpha}\right]\right\}, & x < b \\ 1 & x \geq b \end{cases}$$

83

These three distributions are referred to as the extreme value distributions with types I, II and III known as the Gumbel, Frechet and Weibull families respectively. Parameter 'a' is scale; 'b' is location and α is the shape parameter of the distributions. The question becomes which of the three distributions should be used when analyzing a set of data. The problem is relieved by construction of generalized extreme value distribution

$$f(x;a,b,\alpha)=\frac{1}{\sigma}\left[1+k\left(\frac{x-b}{a}\right)\right]^{-1-1/\alpha}\exp\left\{-\left[1+k\left(\frac{x-b}{a}\right)\right]^{-1/\alpha}\right\}$$

where σ>0, -∞<b, α<∞ and the range of x is such that $[1+\alpha(x-b)/a]>0$. The location, scale and shape of the distribution are controlled by the parameters μ, σ and k respectively.

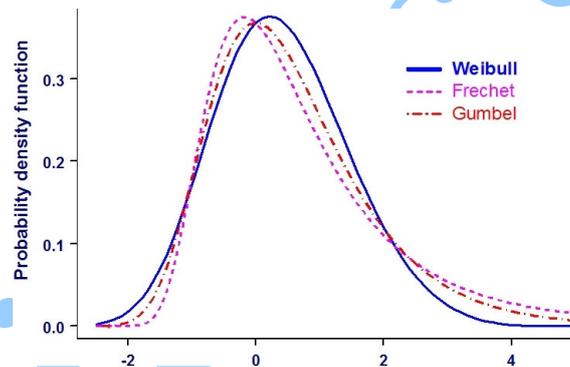

**Figure 1. Probability densities of Weibull, Frechet and Gumbel distribution**

## 2. Maximum Likelihood Method

The maximum likelihood estimation (MLE) technique generally show less bias and provides a more consistent approach to parameter estimation problems. The likelihood of the sample is

$$L=\prod_{i=1}^{n}f(x_i) \quad\quad\ldots\ldots\ldots(1)$$

The logarithm of equation (1), which is log-likelihood (LL), given by

$$LL=\sum_{i=1}^{n}\log[f(x_i)] \quad \ldots\ldots\ldots(2)$$

The simple idea of the maximum likelihood method is to determine the distribution parameters that maximize the likelihood of the given sample. Because the logarithmic is a monotonic function, this is equivalent to maximizing the log-likelihood. Note that f(x) (probability density function) is always less than one so that all terms of the sum for LL are negative, consequently larger log-likelihoods are associated with smaller numerical values [HXN08].

## 3. Selection Criteria for best distribution

The best distribution model is selected on the basis of goodness of fit criteria which is tested by (Anderson-Darling) test. The A-D test procedure is a general test to compare the fit of an observed cumulative distribution function to an expected cumulative distribution function. The test statistic of A-D test is

$$A^2 = -n - \frac{1}{n}\sum_{i=1}^{n}(2i-1)[\ln F(X_i)+\ln\{1-F(X_{n-i+1})\}]$$

where n is the number of observations and F(X) is the Cumulative Density Function (CDF) for the data. Quantile-Quantile (Q-Q) plot and probability difference graph were also used to visual inspection for the selection for the best fitted.

## 4. Return Level

This expression allows us to calculate the maximum likelihood estimates of the parameters. In the most applications the quantiles of the fitted distribution are to interest as they allows us to make predictions about the level exceeded once every 1/(1-p) years on the average. This is called the return level and is given by [NC07]:

$$x_p = \mu - \frac{\sigma}{k}\left[1 - \left\{-\log(1-\frac{1}{P})\right\}^{-k}\right] \quad \ldots (3)$$

## 5. Data source

The data employed in this study are 51 observations of extreme maximum rainfall of year 1956-2006, collected form Birsa Agricultural University, Ranchi, Jharkhand.

## 6. Result and discussion

The study uses 51 years of yearly maximum rainfall data of Ranchi. A quick glance at Fig.-2 will not give much insight to long term change. The summary statistics of extreme rainfall data is illustrated in Table-1, which revealed that data have slight flat tail at right side i.e. positively skewed and platikurtic with high variability. Maximum likelihood estimates of the Gumbel Max, Frechet and Weibull distribution are given in Table-2 and probability density functions (PDF) of Gumbel Max, Frechet, Weibull and GEV distributions are indicated in Fig.-3, respectively.

      In the fitted distributions Gumbel Max, Frechet and Weibull distributions are two parameters distribution and GEV is three parameters distribution, so that in Table:2 Gumbel Max, Frechet and Weibull have only two parameters but GEV has all the three parameters. The Table -2 indicates that two parameter distribution having the comparatively larger value of their parameter than GEV distribution.

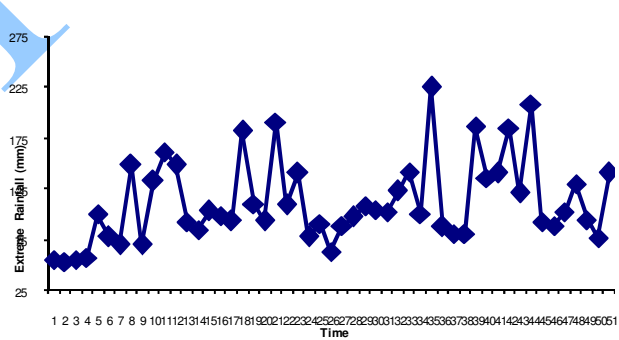

**Figure 2. Fluctuations in extreme rainfall from 1956 to 2006**

**Table 1. Descriptive Statistics of Extreme Rainfall (mm)**

| Statistics | Value |
|---|---|
| Sample Size | 51 |
| Range | 172.30 |
| Mean | 112.09 |
| Variance | 1686.53 |
| Standard Deviation | 41.07 |
| Coefficient of Variation | 36.63 |
| Standard Error | 5.75 |
| Skew ness | 0.926 |
| Kurtosis | 0.397 |

**Table 2. Parameters of fitted Extreme Value Distribution**

| Gumbel Max | | | |
|---|---|---|---|
| **Parameter** | **Location** | **Shape** | **Scale** |
| **Estimate** | 93.61 | 32.02 | --- |
| Frechet | | | |
| **Estimate** | --- | 3.37 | 88.16 |
| Weibull | | | |
| **Estimate** | --- | 3.34 | 122.27 |
| GEV | | | |
| **Estimate** | 92.41 | 0.06 | 30.85 |

## 7. Goodness of Fit test (A-D test)

Goodness of fit for the fitted models is tested by A-D test illustrated in the Table: 3. It is reported from the table that all the models are successfully passed the goodness of fit test (A-D test) at 5% level of significant. That means all the models can easily occupy the uncertainties pattern in the data.

**Table 3. Goodness of Fit test for the fitted model (A-D test)**

| Gumbel Max | | Frechet | | Weibull | | GEV | |
|---|---|---|---|---|---|---|---|
| Statistic | Crit. Val. | Statistic | Crit. Val. | Statistic | Crit. Val. | Statistic | Crit. Val. |
| $0.381^{ns}$ | 2.502 | $0.844^{ns}$ | 2.502 | $1.405^{ns}$ | 2.502 | $0.333^{ns}$ | 2.502 |
| **PASS** | | **PASS** | | **PASS** | | **PASS** | |
| **ns: Non-Significant at 5% level of significant, Crit. Val.: Critical Value** | | | | | | | |

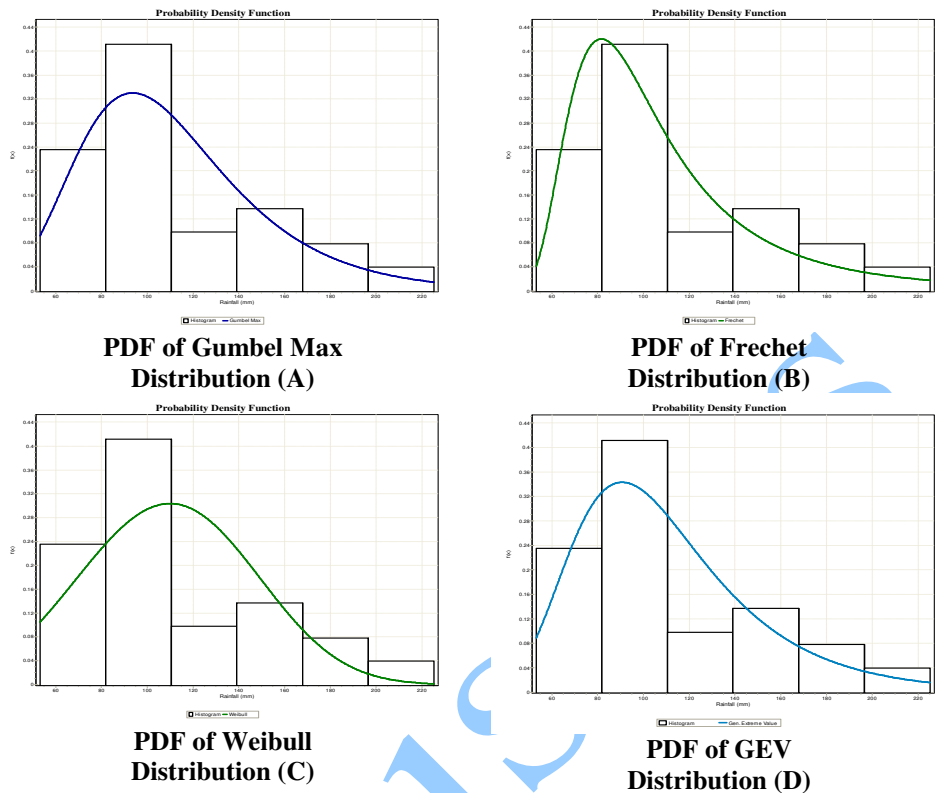

**PDF of Gumbel Max Distribution (A)**

**PDF of Frechet Distribution (B)**

**PDF of Weibull Distribution (C)**

**PDF of GEV Distribution (D)**

**Figure 3. PDF of various distributions for Extreme Maximum Rainfall**

## 8. Normality (Q-Q Plot)

Normality of the fitted distributions is assessed by Q-Q plot indicated in the Figure: 4. It is obvious form the figure that quintiles of GEV distribution (Figure: 4 (D)) mostly occupied the reference line compare to other distributions. In figure: 5 Probability difference chart revealed that inner most (sky color) quintiles are much closer to horizontal reference line belongs to GEV distribution. So, on the basis overall statistical selection criteria GEV is the best fitted distributional model to assess the extreme maximum rainfall in Ranchi.

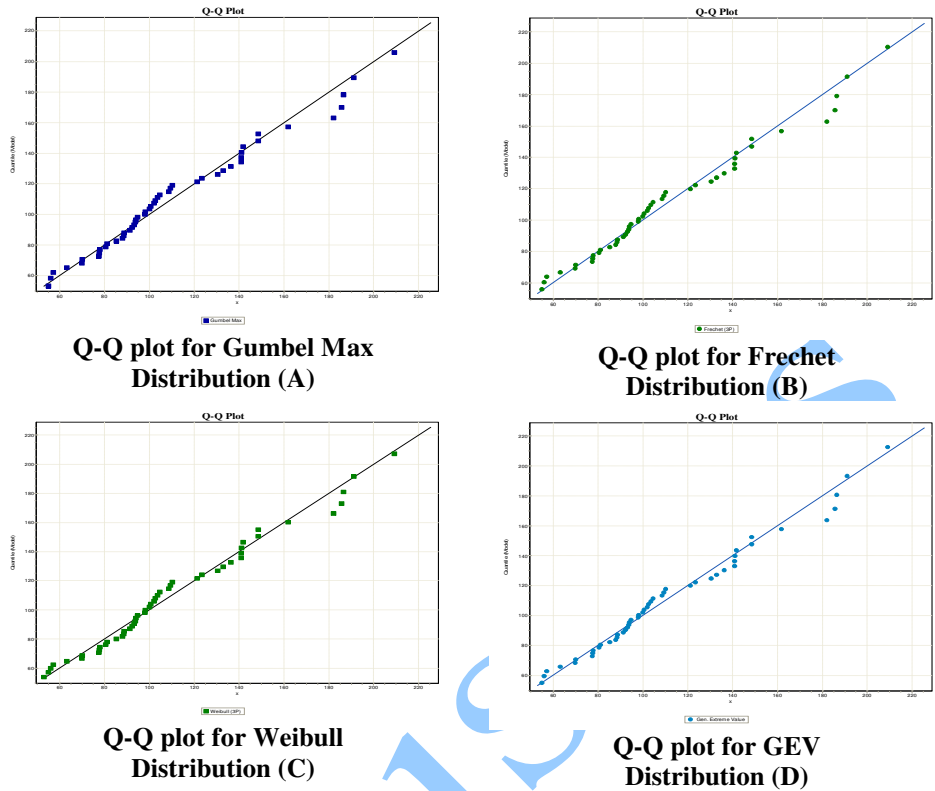

**Q-Q plot for Gumbel Max Distribution (A)**

**Q-Q plot for Frechet Distribution (B)**

**Q-Q plot for Weibull Distribution (C)**

**Q-Q plot for GEV Distribution (D)**

**Figure 4. Q-Q Plot for fitted distributions to assessing normality**

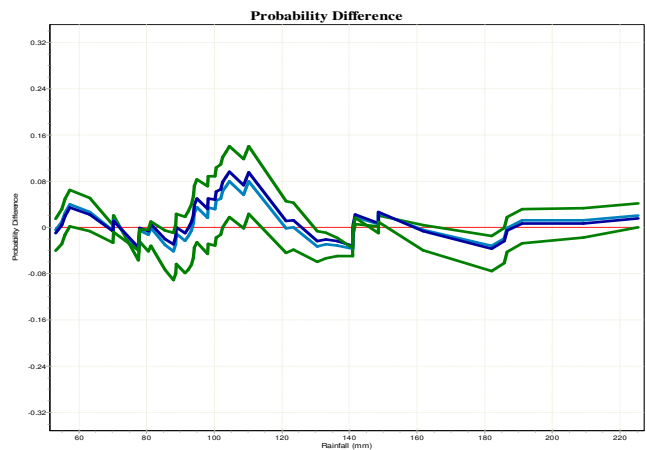

**Figure 5. Probability difference chart for the fitted distribution model**

## 9. Return Level

Once the best model for the data has been selected, the interest is in deriving the return levels of extreme maximum rainfall. The p year return level, say $x_p$, is the level exceeded on average only once in p years already indicated in equation (3). The return level for 5, 10, 50, 100 and 200 years are indicated in the Table -4

Table 4. Return level of maximum extreme rainfall (mm)

| Return Level (mm) | | | | | |
|---|---|---|---|---|---|
| Year | P = 5 | P= 10 | P= 50 | P=100 | P= 200 |
| Rainfall (mm) | 169.67 | 196.94 | 261.39 | 290.61 | 320.98 |

It is revealed by Table 4 that rainfall consistently increasing over the 200 years which is 320.98 mm. This rainfall value is the least maximum extreme value of rainfall which will occur after 200 years.

**Conclusions**

A literature survey shows Extreme Value Theory (EVT) to be a reliable tool for climate extreme scenarios construction and for this purpose maximum likelihood method is used for the evaluation of distribution parameters for extreme rainfall data. Generalized extreme value model is found to be most suitable model with fulfilling all statistical selection criteria then return level for the model is constructed to predict the maximum rainfall for long run in future. It is consistently increasing from time to time for the next 200 years. For further study, researcher can make a long term prediction for other weather parameter which are directly or indirectly affect to the sector like agriculture, production industry and other sectors on which human life is dependent.

**Acknowledgement**